# Making Abstract Domains Condensing

ROBERTO GIACOBAZZI
Università di Verona, Italy
FRANCESCO RANZATO
Università di Padova, Italy
and
FRANCESCA SCOZZARI
Università di Pisa, Italy

In this paper we show that reversible analysis of logic languages by abstract interpretation can be performed without loss of precision by systematically refining abstract domains. The idea is to include semantic structures into abstract domains in such a way that the refined abstract domain becomes rich enough to allow approximate bottom-up and top-down semantics to agree. These domains are known as condensing abstract domains. Substantially, an abstract domain is condensing if goal-driven and goal-independent analyses agree, namely no loss of precision is introduced by approximating queries in a goal-independent analysis. We prove that condensation is an abstract domain property and that the problem of making an abstract domain condensing boils down to the problem of making the domain complete with respect to unification. In a general abstract interpretation setting we show that when concrete domains and operations give rise to quantales, i.e. models of propositional linear logic, objects in a complete refined abstract domain can be explicitly characterized by linear logic-based formulations. This is the case for abstract domains for logic program analysis approximating computed answer substitutions where unification plays the role of multiplicative conjunction in a quantale of idempotent substitutions. Condensing abstract domains can therefore be systematically derived by minimally extending any, generally non-condensing domain, by a simple domain refinement operator.

## 1. INTRODUCTION

Logic program analysis and optimization algorithms are often goal-directed. This means that the analysis is constructively derived from a goal-directed semantics and the properties of the resulting analysis, such as its precision, depend on this choice. For instance, it is well known that, in general, goal-independent analyses of logic programs, like those obtainable with bottom-up/top-down analyzers [Barbuti et al.

Authors' addresses: R. Giacobazzi, Dipartimento di Informatica, Università di Verona, Strada Le Grazie 15, Ca' Vignal 2, 37134 Verona, Italy, e-mail: giaco@sci.univr.it; F. Ranzato, Dipartimento di Matematica Pura ed Applicata, Università di Padova, Via Belzoni 7, 35131 Padova, Italy, e-mail: franz@math.unipd.it; F. Scozzari, Dipartimento di Informatica, Università di Pisa, Corso Italia 40, 56125 Pisa, Italy, e-mail: scozzari@di.unipi.it.

1993; Bruynooghe 1991; Codish et al. 1994], may be less precise than goal-directed ones (cf. [Marriott and Søndergaard 1993, Section 4]). Goal-independent analysis can be thought of as an analysis for all possible initial queries, while in a goal-directed analysis, a given query with a fixed initial state is executed in the abstract domain. In the latter case, the result may be refined by knowing in advance the initial calls. On the contrary, it may also happen that goal-independent analyses are more precise than goal-directed ones, for instance when failure properties are observed.

## 1.1   The problem

The problem of making the analysis independent from the choice of the initial query has been considered by many authors (see e.g. [Codish and Lagoon 2000; Debray 1994; Jacobs and Langen 1992; King and Lu 2002; Langen 1990; Giacobazzi and Scozzari 1998; Marriott and Søndergaard 1993]. The idea of *condensing procedures* introduced by Langen [1990] captures the essence of this discussion, providing a characterization of goal-independent evaluation of procedure calls: The approximation of the semantics of each predicate (called condensed procedure) defined in a program is pre-computed in such a way that any specific call can be approximated without computing a fixpoint, but simply by unifying it against the condensed procedure defining that predicate. However, condensation may loose precision. A loss of precision occurs in condensed procedures when the abstract computation of a procedure call cannot be reconstructed by unifying that call with the corresponding condensed procedure. This is due to the properties of abstract unification on some abstract domains. A domain where no loss of precision occurs in evaluating condensed procedures is called *condensing*. The problem of systematically designing condensing abstract domains is still open. The relevance of this problem relies upon the importance of condensing abstract domains in efficient static program analysis. Moreover, few condensing abstract domains are known, all of them being downward closed domains. In this context, it is highly desirable to have a formal setting where possibly non-downward closed condensing abstract domains can be designed and proved correct, e.g. for relevant program properties like variable aliasing [Jacobs and Langen 1992; Langen 1990].

## 1.2   The main result

In this paper we give a domain-theoretic characterization of condensing abstract interpretations. We prove that it is always possible to make abstract domains condensing by minimally refining domains, i.e. by introducing the least amount of information that makes the domain condensing. We prove that this is an instance of a more general problem of making a domain complete with respect to some semantic operator. The intuition is that a complete abstract interpretation induces an abstract semantics where no loss of precision, relatively to the power of representation of the underlying abstract domains, is accumulated by computing with abstract objects [Cousot and Cousot 1979]. In static program analysis, decidability issues commonly force to sacrifice completeness for achieving termination and/or efficiency; examples of complete abstract interpretations more frequently occur in other fields of application. For instance, several complete abstractions of algebraic polynomial systems have been studied by Cousot and Cousot [1997], and many complete abstract interpretations can be found in comparative program semantics [Cousot 1997; Cousot and Cousot 1992; Giacobazzi 1996] and in model checking by abstract interpretation [Cousot and Cousot 2000; Ranzato 2001].

The possibility of making abstract domains complete with respect to any continuous semantic operator has been shown in [Giacobazzi et al. 2000], where we proved that any abstract domain $A$ can always be constructively extended into the most abstract domain which includes $A$ and is complete for a given continuous function $f$ — the resulting domain is called the *complete shell* of $A$ and $f$. In this paper we apply this technique to systematically derive condensing abstract domains. In particular we consider the problem of minimally modifying abstract domains in order to make them condensing yet providing an easily representable structure for the objects of the refined domain. We prove that this is an instance of a particular completeness problem arising when concrete semantic domains and operations give rise to a particular algebraic structure called *quantale*. Quantales are well-known algebraic structures which turn out to be models of propositional linear logic (see [Rosenthal 1990; Yetter 1990]). This is particularly important in our context, because quantales naturally model a number of different and novel notions of completeness arising in abstract interpretation, including condensation as an instance. Interestingly, in this context it turns out that the objects of complete refined abstract domains can be elegantly represented as linear implications, with a clean logical interpretation. More in detail, a quantale $\langle C_{\leq}, \otimes \rangle$ consists of a complete lattice $C_{\leq}$ together with a binary operation $\otimes : C \times C \longrightarrow C$ which is additive (i.e., preserves arbitrary lub's) on both arguments. As a main feature, quantales support a notion of linear implication between domain's objects: Given $a, b \in C$, there exists a unique greatest object $a \multimap b \in C$ which, when combined by $\otimes$ with $a$, gives a result which is approximated by $b$. In other terms, the following *modus ponens* law $a \otimes x \leq b \iff x \leq a \multimap b$ holds. When refining abstract domains in order to get completeness in a setting where concrete interpretations are quantales, linear implication allows us to elegantly characterize complete domain objects in a variety of situations. It is worth noting that an efficient representation of abstract objects in abstract interpretation is essential in order to automatically (or, at least, quickly) implement abstract domains, e.g. by exploiting the logical properties of the abstract objects (see the use of Binary Decision Diagrams in the implementation of classical propositional logic-based abstract domains for groundness analysis and in abstract model checking); to study the properties of abstract operations, like for instance their precision; to help the intuition to understand how specific abstract domains work. Let us denote by $\mathrm{uco}(C)$ the complete lattice of all abstract domains (modulo isomorphic representation of their objects) abstracting a given domain $C$ (cf. [Cousot and Cousot 1979]). This lattice is ordered by the relative precision of domains: For any $A, B \in \mathrm{uco}(C)$, $A \sqsubseteq B$ if $A$ is more concrete (more precise) than $B$. Given a quantale $\langle C_{\leq}, \otimes \rangle$ and an abstract domain $A \in \mathrm{uco}(C)$, we characterize the most abstract domain $X \in \mathrm{uco}(C)$ such that $X \sqsubseteq A$ and $X$ is complete for $\otimes$, namely if $\alpha_X : C \longrightarrow X$ is the corresponding abstraction map then the equation $\alpha_X(\cdot \otimes \cdot) = \alpha_X(\alpha_X(\cdot) \otimes \alpha_X(\cdot))$ holds. We prove that a domain is condensing if and only if a weakened form of completeness holds: $\alpha_X(\alpha_X(\cdot) \otimes \alpha_X(\cdot)) = \alpha_X(\cdot \otimes \alpha_X(\cdot)) = \alpha_X(\alpha_X(\cdot) \otimes \cdot)$. Thus, given $A \in \mathrm{uco}(C)$, we characterize the most abstract domain $X \in \mathrm{uco}(C)$ such that $X \sqsubseteq A$ and $X$ is condensing. Intuitively, $X$ is condensing when there is no loss of precision in observing in $X$ the result of $\otimes$ when one of its arguments is approximated in $X$. The set of idempotent substitutions endowed with unification forms a quantale. As a consequence abstract domains can be refined and constructively made condensing, providing their objects with an elegant logical characterization as linear implications in the quantale of idempotent substitutions. This is a generalization of an analogous result given by Giacobazzi and Scozzari [1998], which characterizes condensing *downward closed* domains as solutions of domain equations involving intuitionistic implications.

## 2. BASIC NOTIONS

### 2.1 Notation

If $S$ and $T$ are sets, then $\wp(S)$ denotes the powerset of $S$, $S \longrightarrow T$ denotes the set of all functions from $S$ to $T$, and for a function $f : S \longrightarrow T$ and $X \subseteq S$, $f(X) \overset{\text{def}}{=} \{f(x) \mid x \in X\}$. By $g \circ f$ we denote the composition of the functions $f$ and $g$, i.e., $g \circ f \overset{\text{def}}{=} \lambda x.g(f(x))$. The identity function $\lambda x.x$ is denoted $id$. The notation $P_\leq$ denotes a poset $P$ with ordering relation $\leq$, while $\langle C, \leq, \vee, \wedge, \top, \bot \rangle$ denotes a complete lattice $C$, with ordering $\leq$, lub $\vee$, glb $\wedge$, greatest element (top) $\top$, and least element (bottom) $\bot$. Somewhere, $\leq_P$ will be used to denote the underlying ordering of a poset $P$, and $\vee_C$, $\wedge_C$, $\top_C$ and $\bot_C$ will denote operations and elements of a complete lattice $C$. Let $P$ be a poset and $S \subseteq P$. Then, $max(S) \overset{\text{def}}{=} \{x \in S \mid \forall y \in S.\ x \leq_P y \Rightarrow x = y\}$ denotes the set of maximal elements of $S$ in $P$; also, the downward closure of $S$ is defined by $\downarrow S \overset{\text{def}}{=} \{x \in P \mid \exists y \in S.\ x \leq_P y\}$, and for $x \in P$, $\downarrow x$ is a shorthand for $\downarrow \{x\}$. We use the symbol $\sqsubseteq$ to denote pointwise ordering between functions: If $S$ is any set, $P$ a poset, and $f, g : S \longrightarrow P$ then $f \sqsubseteq g$ if for all $x \in S$, $f(x) \leq_P g(x)$. Let $C$ and $D$ be complete lattices. Then, $C \overset{\text{m}}{\longrightarrow} D$, $C \overset{\text{c}}{\longrightarrow} D$, $C \overset{\text{a}}{\longrightarrow} D$, and $C \overset{\text{coa}}{\longrightarrow} D$ denote, respectively, the set of all monotone, (Scott-)continuous, additive, and co-additive functions from $C$ to $D$. Recall that $f \in C \overset{\text{c}}{\longrightarrow} D$ iff $f$ preserves lub's of (non-empty) chains, and $f : C \longrightarrow D$ is (completely) additive if $f$ preserves lub's of arbitrary subsets of $C$ (emptyset included). Co-additivity is dually defined. We denote by $lfp(f)$ and $gfp(f)$, respectively, the least and greatest fixpoint, when they exist, of an operator $f$ on a poset. If $f \in C \overset{\text{c}}{\longrightarrow} C$ then $lfp(f) = \vee_{i \in \mathbb{N}} f^i(\bot_C)$, where, inductively, $f^0(x) \overset{\text{def}}{=} x$ and $f^{i+1}(x) \overset{\text{def}}{=} f(f^i(x))$. Dually, if $f$ is co-continuous then $gfp(f) = \wedge_{i \in \mathbb{N}} f^i(\top_C)$. $\{f^i(\bot_C)\}_{i \in \mathbb{N}}$ and $\{f^i(\top_C)\}_{i \in \mathbb{N}}$ are called, respectively, the upper and lower Kleene's iteration sequences of $f$.

### 2.2 Logic programming

Let $\mathcal{V}$ be an infinite set of variables and $\texttt{Term}$ be the set of terms with variables in $\mathcal{V}$. A substitution $\sigma$ is a mapping from $\mathcal{V}$ to $\texttt{Term}$ such that $\{v \in \mathcal{V} \mid \sigma(v) \neq v\}$ is a finite set. By $s\sigma$ and $\sigma(s)$ we denote the application of $\sigma$ to any syntactic object $s$, while $vars(s)$ denotes the set of variables occurring in $s$. A term $t$ is ground if $vars(t) = \varnothing$. The composition of substitutions is denoted by $\sigma \circ \theta = \lambda x.\sigma(\theta(x))$. The set of *idempotent* substitutions modulo renaming $\sim$ (i.e., given $\theta$ and $\sigma$ idempotent, $\theta \sim \sigma$ if and only if there exist two substitutions $\beta$ and $\delta$ such that $\theta = \beta \circ \sigma$ and $\sigma = \delta \circ \theta$) is denoted by $\texttt{Sub}$. $\texttt{Sub}$ is partially ordered by instantiation, denoted by $\preceq$:

$\sigma \preceq \theta$ iff $\exists \delta \in \texttt{Sub}. \ \sigma = \delta \circ \theta$. By adding to $\texttt{Sub}$ an extra object $\tau$ as least element, one gets a complete lattice $\langle \texttt{Sub}^\tau, \preceq, \vee, \wedge, \epsilon, \tau \rangle$, where $\vee$ is the least general anti-instance, $\epsilon$ is the empty substitution, and $\wedge$ is the standard unification, which is unique modulo renaming (see [Eder 1985] and [Palamidessi 1990, Sections 3 and 4] for the details). In the following, for $\sigma, \theta \in \texttt{Sub}$, we will write $\sigma \wedge \theta \neq \tau$ to denote that $\sigma$ and $\theta$ unify.

### 2.3   The lattice of abstract interpretations

In standard Cousot and Cousot's abstract interpretation theory, abstract domains can be equivalently specified either by Galois connections (GCs), i.e., adjunctions, or by upper closure operators (uco's) [Cousot and Cousot 1979]. In the first case, concrete and abstract domains $C$ and $A$ (both assumed to be complete lattices) are related by a pair of adjoint functions of a GC $(\alpha, C, A, \gamma)$, where $\alpha$ and $\gamma$ are the abstraction and concretization maps. It is usually assumed that $(\alpha, C, A, \gamma)$ is a Galois insertion (GI), i.e., $\alpha$ is onto or, equivalently, $\gamma$ is 1-1. In the second case, instead, an abstract domain is specified as a uco on the concrete domain $C$, i.e., a monotone, idempotent and extensive operator on $C$. These two approaches are perfectly equivalent, modulo isomorphic representation of domain's objects. Given a complete lattice $C$, it is well known that the set $\mathrm{uco}(C)$ of all uco's on $C$, endowed with the pointwise ordering $\sqsubseteq$, gives rise to the complete lattice $\langle \mathrm{uco}(C), \sqsubseteq, \sqcup, \sqcap, \lambda x. \top_C, id \rangle$. Let us recall that each $\rho \in \mathrm{uco}(C)$ is uniquely determined by the set of its fixpoints, which is its image, i.e. $\rho(C) = \{ x \in C \mid \rho(x) = x \}$, since $\rho = \lambda x. \wedge \{ y \in C \mid y \in \rho(C), x \leq y \}$. Moreover, a subset $X \subseteq C$ is the set of fixpoints of a uco on $C$ iff $X$ is meet-closed, i.e. $X = \mathcal{M}(X) \stackrel{\mathrm{def}}{=} \{ \wedge Y \mid Y \subseteq X \}$ (note that $\top_C = \wedge \varnothing \in \mathcal{M}(X)$). For any $X \subseteq C$, $\mathcal{M}(X)$ is called the Moore-closure of $X$, and $X$ is a generator set for $\mathcal{M}(X)$. Also, $\rho \sqsubseteq \eta$ iff $\eta(C) \subseteq \rho(C)$; in this case, $\rho$ is a so-called refinement of $\eta$, and if $\rho \sqsubseteq \eta$ then $\rho \circ \eta = \eta \circ \rho = \eta$. Often, we will identify closures with their sets of fixpoints. This does not give rise to ambiguity, since one can distinguish their use as functions or sets according to the context. In view of the equivalence above, throughout the paper, $\langle \mathrm{uco}(C), \sqsubseteq \rangle$ will play the role of the lattice of abstract interpretations of $C$ [Cousot and Cousot 1977; Cousot and Cousot 1979], i.e. the complete lattice of all the abstract domains of the concrete domain $C$. When an abstract domain $A$ is specified by a GI $(\alpha, C, A, \gamma)$, $\rho_A \stackrel{\mathrm{def}}{=} \gamma \circ \alpha \in \mathrm{uco}(C)$ is the corresponding uco on $C$. The ordering on $\mathrm{uco}(C)$ corresponds to the standard order used to compare abstract domains with regard to their precision: $A_1$ is more precise than $A_2$ (i.e., $A_1$ is more concrete than $A_2$ or $A_2$ is more abstract than $A_1$) iff $A_1 \sqsubseteq A_2$ in $\mathrm{uco}(C)$. Lub and glb on $\mathrm{uco}(C)$ have therefore the following reading as operators on domains. Let $\{ A_i \}_{i \in I} \subseteq \mathrm{uco}(C)$: (i) $\sqcup_{i \in I} A_i$ is the most concrete among the domains which are abstractions of all the $A_i$'s; (ii) $\sqcap_{i \in I} A_i$ is the most abstract among the domains which are more concrete than every $A_i$ – this domain is also known as reduced product of all the $A_i$'s.

### 2.4   Completeness in abstract interpretation

Completeness in abstract interpretation uniquely depends upon the abstraction map [Giacobazzi and Ranzato 1997]. Let us consider the simple case of an abstract interpretation specified by an abstract domain $A$ and an abstract operation $f^\sharp : A \longrightarrow A$ approximating a concrete semantic operation $f : C \longrightarrow C$. Then, $f^\sharp$ is (sound and) complete if $\rho \circ f = f^\sharp \circ \rho$, where $\rho \in uco(C)$ is the uco associated with $A$. It turns out that if $f^\sharp$ is complete then the best correct approximation of $f$ in $A$, i.e. $\rho \circ f : A \longrightarrow A$, is complete as well, and, in this case, $f^\sharp$ indeed coincides with $\rho \circ f$. Thus, for any $A$, one can define a complete abstract semantic operation $f^\sharp : A \longrightarrow A$ over $A$ if and only if $\rho \circ f : A \longrightarrow A$ is complete. Hence, an abstract domain $\rho \in uco(C)$ is defined to be complete for $f$ iff $\rho \circ f = \rho \circ f \circ \rho$ holds. This simple observation makes completeness an abstract domain property, namely an intrinsic characteristic of the abstract domain. It is also worth recalling that, by a well-known result [Cousot and Cousot 1979, Theorem 7.1.0.4], complete abstract domains are "fixpoint complete" as well, i.e., if $\rho$ is complete for $f$ then $\rho(lfp(f)) = lfp(\rho \circ f)$, while the converse, in general, does not hold.

In [Giacobazzi et al. 2000] we gave a constructive characterization of complete abstract domains, under the assumption of dealing with Scott-continuous concrete functions. This result allows us to systematically derive complete abstract domains from non-complete ones in a minimal way. The idea is to build the greatest (i.e., most abstract) domain in $uco(C)$ which includes a given domain $A$ and which is complete for a set $F$ of (continuous) functions, i.e., for each function in $F$. Given a set of continuous functions $F \subseteq C \xrightarrow{c} C$, Giacobazzi et al. [2000] define a mapping $\mathcal{R}_F : uco(C) \longrightarrow uco(C)$ as follows:

$$\mathcal{R}_F(\rho) \stackrel{\text{def}}{=} \mathcal{M}(\bigcup_{f \in F, a \in \rho} max(\{x \in C \mid f(x) \le a\})).$$

THEOREM 2.1. [Giacobazzi et al. 2000] *A domain* $\rho \in uco(C)$ *is complete for* $F$ *iff* $\rho \sqsubseteq \mathcal{R}_F(\rho)$. *Moreover,* $\mathcal{R}_F$ *is co-additive.*

Thus, the most abstract domain which includes $\rho$ and which is complete for $F$ is $gfp(\lambda\eta.\rho \sqcap \mathcal{R}_F(\eta))$. This domain is called the *complete shell* of $\rho$ for $F$ (see [Giacobazzi et al. 2000] for more details).

## 2.5 Quantales and linear logic

Quantales originated in the algebraic foundations of the so-called *quantum logic*. Afterwards, they have been successfully considered as algebraic models of Girard's linear logic [Rosenthal 1990; Yetter 1990]. Informally, quantales can be thought of as a generalization of Boolean algebras, where the modus ponens law $a \wedge (a \Rightarrow b) \le b$ holds relatively to a binary operation $\otimes$ of "conjunction" possibly different from the meet. The basic idea in a quantale is to guarantee that, for any two objects $a$ and $b$, there exists a greatest (i.e., most abstract in abstract interpretation terms) object $c$ such that $a \otimes c \le b$. In the following, we restrict our attention to commutative quantales, i.e., quantales where the binary operation $\otimes$ is commutative. More formally, a (commutative) quantale is an algebra $\langle C_{\le}, \otimes \rangle$ such that:

—$\langle C, \le, \vee, \wedge, \top, \bot \rangle$ is a complete lattice;

—$\otimes : C \times C \longrightarrow C$ is a commutative and associative operation, i.e., $a \otimes b = b \otimes a$ and $(a \otimes b) \otimes c = a \otimes (b \otimes c)$, for any $a, b, c \in C$;

—$a \otimes (\bigvee_{i \in I} b_i) = \bigvee_{i \in I}(a \otimes b_i)$, for any $a \in C$ and $\{b_i\}_{i \in I} \subseteq C$.

In other words, a quantale is a complete lattice endowed with a commutative and associative "product" $\otimes$ which distributes over arbitrary lub's. Common ex-amples of quantales are complete Boolean algebras, which become quantales by considering as $\otimes$ their meet operation. In particular, for any set $A$, the algebra $\langle \wp(A)_{\subseteq}, \bigcap \rangle$ is a quantale. Also, given a commutative and associative operation $\cdot : A \times A \longrightarrow A$, a further basic example of quantale is $\langle \wp(A)_{\subseteq}, \otimes \rangle$, where $X \otimes Y \stackrel{\text{def}}{=} \bigcup \{x \cdot y \mid x \in X, y \in Y\}$ is the lifting of the operation $\cdot$ to sets.

The fundamental property of quantales is that, for any $a \in C$, the function $\lambda x. a \otimes x$ has a right adjoint, denoted by $\lambda x. a \multimap x$. This is equivalent to say that one can define a binary operation $\multimap : C \times C \longrightarrow C$ such that, for all $a, b, c \in C$, the following property holds:

$$a \otimes b \leq c \iff b \leq a \multimap c.$$

This is a straight consequence of the fact that, for all $a \in C$, $\lambda x. a \otimes x$ is additive, and therefore, it has a unique right adjoint $\lambda x. a \multimap x$ giving rise to a GC. This right adjoint $\multimap : C \times C \longrightarrow C$ is therefore defined as follows:

$$a \multimap c \stackrel{\text{def}}{=} \bigvee \{b \in C \mid a \otimes b \leq c\}.$$

A quantale $\langle C_{\leq}, \otimes \rangle$ is called *unital* if there exists an object $\mathbf{1} \in C$, called unit, such that $\mathbf{1} \otimes a = a = a \otimes \mathbf{1}$, for all $a \in C$. $\langle \wp(A)_{\subseteq}, \bigcap \rangle$ is a trivial example of unital, commutative quantale, where $A$ is the unit.

From a logical point of view, it is well known that quantales turn out to be models of (commutative) linear logic [Rosenthal 1990; Yetter 1990], where the linear implication is interpreted as the operation $\multimap$. The next proposition summarizes the basic properties of linear implication (see [Rosenthal 1990]).

PROPOSITION 2.2. *Let $\langle C_{\leq}, \otimes \rangle$ be a unital, commutative quantale with unit $\mathbf{1}$, $\{x_i\}_{i \in I} \subseteq C$ and $a, b, c \in C$.*

   (i)   $a \otimes (a \multimap c) \leq c$                           (ii)   $a \multimap (b \multimap c) = (b \otimes a) \multimap c$

 (iii)   $a \multimap (\bigwedge_{i \in I} x_i) = \bigwedge_{i \in I}(a \multimap x_i)$         (iv)   $(\bigvee_{i \in I} x_i) \multimap c = \bigwedge_{i \in I}(x_i \multimap c)$

  (v)   $a \multimap (b \multimap c) = b \multimap (a \multimap c)$       (vi)   $\mathbf{1} \multimap a = a$

 (vii)   $c \leq (c \multimap a) \multimap a$                  (viii)   $((c \multimap a) \multimap a) \multimap a = c \multimap a$

 (ix)   *if  $b \leq c$  then  $a \otimes b \leq a \otimes c$*

In particular, from the above properties, it is easy to check that for all $a \in C$, $\lambda x. (x \multimap a) \multimap a \in \text{uco}(C)$.

## 3. COMPLETENESS IN LOGICAL FORM

In this section we consider completeness in quantales, providing a linear logic-based characterization of complete abstract interpretations of quantales. Let $\langle C_{\leq}, \otimes \rangle$ be a unital, commutative quantale playing the role of concrete interpretation, that is, $C$ is the concrete domain provided with a semantic operation $\otimes : C \times C \longrightarrow C$. Let $\rho \in \text{uco}(C)$ be an abstract domain. Recall that $\rho$ is complete for $\otimes$ when for all concrete objects $x, y \in C$, $\rho(\rho(x) \otimes \rho(y)) = \rho(x \otimes y)$. This is more compactly denoted by the equation $\rho \circ \otimes \circ \langle \rho, \rho \rangle = \rho \circ \otimes$. Given any $\eta \in \text{uco}(C)$, we define the following set of unary (additive) functions $F_\eta \subseteq C \xrightarrow{\text{a}} C$:

$$F_\eta \stackrel{\text{def}}{=} \{\lambda x. x \otimes y \mid y \in \eta\}.$$

In particular, $F_{id}$ will be also denoted by $F_C$. It turns out that completeness of $\rho$ for $\otimes$ is equivalent to completeness of $\rho$ for $F_C$.

LEMMA 3.1. *Let $\langle C_\leq, \otimes \rangle$ be a commutative quantale and $\rho \in \mathrm{uco}(C)$. The following are equivalent.*

(i) *$\rho$ is complete for $\otimes$;*
(ii) *$\rho \circ \otimes \circ \langle \rho, id \rangle = \rho \circ \otimes$;*
(iii) *$\rho$ is complete for $F_C$.*

PROOF. We first show (i) ⇔ (ii). Assume that $\rho \circ \otimes \circ \langle \rho, \rho \rangle = \rho \circ \otimes$. Then, by monotonicity and extensivity of $\rho$, we get $\rho \circ \otimes \leq \rho \circ \otimes \circ \langle \rho, id \rangle \leq \rho \circ \otimes \circ \langle \rho, \rho \rangle = \rho \circ \otimes$. On the other hand, assume that $\rho \circ \otimes \circ \langle \rho, id \rangle = \rho \circ \otimes$. By monotonicity and extensivity of $\rho$, $\rho \circ \otimes \leq \rho \circ \otimes \circ \langle \rho, \rho \rangle = \rho \circ \otimes \circ \langle \rho, id \rangle \circ \langle id, \rho \rangle = \rho \circ \otimes \circ \langle id, \rho \rangle =$ (by commutativity of $\otimes$) $= \rho \circ \otimes \circ \langle \rho, id \rangle = \rho \circ \otimes$.
Thus, $\rho$ is complete iff $\forall x, y \in C$, $\rho(\rho(x) \otimes y) = \rho(x \otimes y)$, and this is equivalent to state that $\rho$ is complete for the set of unary functions $F_C = \{\lambda x. x \otimes y \mid y \in C\}$, which concludes the proof. ☐

COROLLARY 3.2. *Let $\langle C_\leq, \otimes \rangle$ be a commutative quantale and $\rho \in \mathrm{uco}(C)$. The complete shell of $\rho$ for $\otimes$ is $gfp(\lambda \eta. \rho \sqcap \mathcal{R}_{F_C}(\eta))$.*

PROOF. By Lemma 3.1, the complete shell of $\rho$ for $\otimes$ coincides with the complete shell of $\rho$ for $F_C$. Each function in $F_C$ is additive, and therefore continuous. Thus, by applying Theorem 2.1, the complete shell of $\rho$ for $\otimes$ is $gfp(\lambda \eta. \rho \sqcap \mathcal{R}_{F_C}(\eta))$. ☐

Thus, the complete shell of any domain $\rho$ for $\otimes$ can be constructively obtained by iterating the operator $\mathcal{R}_{F_C}$. Our main aim is to show that this operator and, more generally, the family of operators $\mathcal{R}_{F_\eta}$, for any $\eta \in \mathrm{uco}(C)$, can all be characterized in terms of sets of linear implications. Let us define a domain operator $\stackrel{\wedge}{\multimap} : \mathrm{uco}(C) \times \mathrm{uco}(C) \longrightarrow \mathrm{uco}(C)$ by lifting linear implication $\multimap$ to abstract domains as follows: For any $A, B \in \mathrm{uco}(C)$:

$$A \stackrel{\wedge}{\multimap} B \stackrel{\mathrm{def}}{=} \mathcal{M}(\{a \multimap b \in C \mid a \in A, b \in B\}).$$

Hence, $A \stackrel{\wedge}{\multimap} B$ is defined to be the most abstract domain in $\mathrm{uco}(C)$ containing all the linear implications from $A$ to $B$.

THEOREM 3.3. *Let $\langle C_\leq, \otimes \rangle$ be a unital, commutative quantale. For any $\rho, \eta \in \mathrm{uco}(C)$, $\mathcal{R}_{F_\eta}(\rho) = \eta \stackrel{\wedge}{\multimap} \rho$.*

PROOF. Let us prove that $\mathcal{R}_{F_\eta}(\rho) = \mathcal{M}(\{y \multimap a \mid y \in \eta, a \in \rho\})$.

$$
\begin{aligned}
\mathcal{R}_{F_\eta}(\rho) = && [ \text{ by definition of } \mathcal{R}_{F_\eta} ] \\
\mathcal{M}(\cup_{f \in F_\eta, a \in \rho} max(\{x \in C \mid f(x) \leq a\})) = && [ \text{ by definition of } F_\eta ] \\
\mathcal{M}(\cup_{y \in \eta, a \in \rho} max(\{x \in C \mid x \otimes y \leq a\})) = && [ \text{ by commutativity of } \otimes ] \\
\mathcal{M}(\cup_{y \in \eta, a \in \rho} max(\{x \in C \mid y \otimes x \leq a\})) = && [ \text{ by definition of } \multimap ] \\
\mathcal{M}(\cup_{y \in \eta, a \in \rho} max(\{x \in C \mid x \leq y \multimap a\})) = && \\
\mathcal{M}(\cup_{y \in \eta, a \in \rho} \{y \multimap a\}) = && \\
\eta \stackrel{\wedge}{\multimap} \rho. &&
\end{aligned}
$$

This closes the proof. ☐

The following basic properties of $\overset{\wedge}{\multimap}$ follow directly from the corresponding properties of the linear implication in quantales.

PROPOSITION 3.4. *For all $A \in \mathrm{uco}(C)$ and $\{B_i\}_{i \in I} \subseteq \mathrm{uco}(C)$, we have:*

(i) $A \overset{\wedge}{\multimap} (\bigsqcap_{i \in I} B_i) = \bigsqcap_{i \in I}(A \overset{\wedge}{\multimap} B_i)$;

(ii) $A \overset{\wedge}{\multimap} \top_{\mathrm{uco}(C)} = \top_{\mathrm{uco}(C)}$;

(iii) $C \overset{\wedge}{\multimap} A \sqsubseteq A$;

(iv) $C \overset{\wedge}{\multimap} A = C \overset{\wedge}{\multimap} (C \overset{\wedge}{\multimap} A)$.

PROOF. Points (i) and (ii) are straightforward.

(iii). By Prop. 2.2 (vi), for all $a \in A$ it holds $\mathbf{1} \multimap a = a$. Since $\mathbf{1} \in C$, it follows that $a = \mathbf{1} \multimap a \in C \overset{\wedge}{\multimap} A$, and therefore $C \overset{\wedge}{\multimap} A \sqsubseteq A$.

(iv). By point (iii), $C \overset{\wedge}{\multimap} A \sqsubseteq A$, and therefore, by right monotonicity of $\overset{\wedge}{\multimap}$, $C \overset{\wedge}{\multimap} (C \overset{\wedge}{\multimap} A) \sqsubseteq C \overset{\wedge}{\multimap} A$. For the other inequality, consider an element belonging to $\{c \multimap a \in C \mid c \in C, a \in C \overset{\wedge}{\multimap} A\}$. By definition, such an element can be written as follows: $c \multimap \bigwedge_{i \in I}(d_i \multimap a_i)$, for suitable $c, d_i \in C$, and $a_i \in A$, for all $i \in I$, where $I$ is a suitable set of indexes. Then,

$$c \multimap \bigwedge_{i \in I}(d_i \multimap a_i) = \bigwedge_{i \in I}(c \multimap (d_i \multimap a_i)) \quad [\text{ by Prop. 2.2 (iii) }]$$
$$= \bigwedge_{i \in I}((d_i \otimes c) \multimap a_i) \quad [\text{ by Prop. 2.2 (ii) }]$$

Since, for all $i \in I$, $d_i \otimes c \in C$ and $a_i \in A$, $(d_i \otimes c) \multimap a_i \in C \overset{\wedge}{\multimap} A$. Then, by monotonicity of the Moore-closure, we get $C \overset{\wedge}{\multimap} A \sqsubseteq C \overset{\wedge}{\multimap} (C \overset{\wedge}{\multimap} A)$.

This concludes the proof. □

It is worth noting that, by points (iii) and (iv) above, the monotone operator $\lambda X. C \overset{\wedge}{\multimap} X : \mathrm{uco}(C) \longrightarrow \mathrm{uco}(C)$ is reductive and idempotent, and therefore it is a lower closure operator on $\mathrm{uco}(C)$. Also, it is important to note that in general $A$ and $A \overset{\wedge}{\multimap} A$ are incomparable abstract domains.

The following result shows that the complete shell of an abstract domain $A$ for $\otimes$ is given by all the linear implications from the concrete domain to $A$. This provides a first representation result for objects of complete abstractions of quantales.

THEOREM 3.5. *Let $\langle C_\leq, \otimes \rangle$ be a unital, commutative quantale and $A \in \mathrm{uco}(C)$. The complete shell of $A$ for $\otimes$ is $C \overset{\wedge}{\multimap} A$.*

PROOF. By Corollary 3.2, the complete shell of $A$ for $\otimes$ is $gfp(\lambda X. A \sqcap \mathcal{R}_{F_C}(X))$, and, by Theorem 3.3, this is $gfp(\lambda X. A \sqcap C \overset{\wedge}{\multimap} X)$. We show that $gfp(\lambda X. A \sqcap C \overset{\wedge}{\multimap} X) = C \overset{\wedge}{\multimap} A$ by computing the corresponding Kleene's iteration sequence.

$$
\begin{aligned}
(\lambda X. A \sqcap C \overset{\wedge}{\multimap} X)(\top_{\mathrm{uco}(C)}) &= A \sqcap C \overset{\wedge}{\multimap} \top_{\mathrm{uco}(C)} &&= A \sqcap \top_{\mathrm{uco}(C)} = A \\
&&&[\text{by Prop. 3.4 (ii)}] \\
(\lambda X. A \sqcap C \overset{\wedge}{\multimap} X)(A) &= A \sqcap C \overset{\wedge}{\multimap} A &&= C \overset{\wedge}{\multimap} A \\
&&&[\text{by Prop. 3.4 (iii)}] \\
(\lambda X. A \sqcap C \overset{\wedge}{\multimap} X)(C \overset{\wedge}{\multimap} A) &= A \sqcap C \overset{\wedge}{\multimap}(C \overset{\wedge}{\multimap} A) &&= A \sqcap C \overset{\wedge}{\multimap} A \\
&&&[\text{by Prop. 3.4 (iv)}] \\
&&&= C \overset{\wedge}{\multimap} A \\
&&&[\text{by Prop. 3.4 (iii)}]
\end{aligned}
$$

Thus, $C \overset{\wedge}{\multimap} A$ actually is the greatest fixpoint of $\lambda X. A \sqcap C \overset{\wedge}{\multimap} X$. □

The relevance of this result stems from the fact that, in the considered case of concrete quantales, the fixpoint construction of the complete shell of an abstract domain converges in two steps, and this provides a clean logical characterization for the objects of the complete shell in terms of linear implications. Furthermore, the following result yields an explicit logical characterization for the abstraction map associated with that complete shell.

THEOREM 3.6. *Let $\langle C_{\leq}, \otimes \rangle$ be a unital, commutative quantale and $A \in \mathrm{uco}(C)$. Let $\rho \in \mathrm{uco}(C)$ be the uco associated with $C \stackrel{\wedge}{\multimap} A$. Then, for all $c \in C$,*

$$\rho(c) = \bigwedge_{a \in A} (c \multimap a) \multimap a.$$

PROOF. Since $A = \prod_{a \in A} \{\top_C, a\}$, by Proposition 3.4 (i), we have that $C \stackrel{\wedge}{\multimap} A = \prod_{a \in A} C \stackrel{\wedge}{\multimap} \{\top_C, a\}$. Let us show that the closure operator $\rho_a \in \mathrm{uco}(C)$ associated with $C \stackrel{\wedge}{\multimap} \{\top_C, a\}$ is $\rho_a = \lambda c.(c \multimap a) \multimap a$, i.e., by Theorem 3.5, $\rho_a$ is the complete shell of $\{\top_C, a\}$ for $\otimes$. Then, the thesis is a straight consequence, since, by definition, for any $c \in C$, $\rho(c) = \bigwedge_{a \in A} \rho_a(c) = \bigwedge_{a \in A}(c \multimap a) \multimap a$. We first show that $\rho_a$ is complete for $\otimes$. By Lemma 3.1, it is enough to show that for any $x, y \in C$, $\rho_a(\rho_a(x) \otimes y) = \rho_a(x \otimes y)$. We prove that $\rho_a(x) \otimes y \leq \rho_a(x \otimes y)$, since this implies $\rho_a(\rho_a(x) \otimes y) \leq \rho_a(x \otimes y)$ and the other inequality always holds. We have that $y \otimes (y \multimap (x \multimap a)) \leq x \multimap a$, and therefore $y \otimes (y \multimap (x \multimap a)) \otimes ((x \multimap a) \multimap a) \leq (x \multimap a) \otimes ((x \multimap a) \multimap a) \leq a$. As a consequence, we have the following inequalities:

$$y \otimes (y \multimap (x \multimap a)) \otimes ((x \multimap a) \multimap a) \leq a$$
$$y \otimes ((x \otimes y) \multimap a) \otimes ((x \multimap a) \multimap a) \leq a$$
$$y \otimes ((x \multimap a) \multimap a) \leq ((x \otimes y) \multimap a) \multimap a$$
$$y \otimes \rho_a(x) \leq \rho_a(x \otimes y).$$

Thus, $\rho_a$ is complete for $\otimes$. Then, in order to conclude, we prove that $\rho_a$ is the greatest domain complete for $\otimes$ which contains the object $a$. Suppose, by contradiction, that there exists $\eta \in \mathrm{uco}(C)$ such that $\eta(a) = a$, $\eta$ is complete for $\otimes$ and $\rho_a \sqsubset \eta$. Therefore, there exists $c \in C$ such that $\eta(c) > \rho_a(c)$, that is $\eta(c) > (c \multimap a) \multimap a$. Then, $\eta(c) \otimes (c \multimap a) \not\leq a$, otherwise we would get $\eta(c) \leq (c \multimap a) \multimap a$, which is a contradiction. As a consequence, $\eta(\eta(c) \otimes \eta(c \multimap a)) \not\leq a$. But, by completeness of $\eta$, $\eta(\eta(c) \otimes \eta(c \multimap a)) = \eta(c \otimes (c \multimap a)) \leq \eta(a) = a$, and this is the contradiction which closes the proof. □

## 4.  CHARACTERIZING CONDENSING ABSTRACT DOMAINS

In this section we give a characterization of condensing abstract domains as solutions of simple abstract domain equations, where the objects of condensing abstract domains have an immediate interpretation in a fragment of propositional linear logic. We consider a core logic programming language computing substitutions. Our basic semantic structure is the unital, commutative quantale $\langle \wp(\mathtt{Sub})_{\subseteq}, \otimes \rangle$, where $\langle \wp(\mathtt{Sub}), \subseteq \rangle$ is a complete lattice and $\otimes : \wp(\mathtt{Sub}) \times \wp(\mathtt{Sub}) \longrightarrow \wp(\mathtt{Sub})$ is the standard lifting of unification $\wedge$ to sets of substitutions, namely:

$$X \otimes Y \stackrel{\mathrm{def}}{=} \{x \wedge y \mid x \in X, \ y \in Y, \ x \wedge y \neq \tau\}.$$

Obviously, $\langle \wp(\mathtt{Sub})_{\subseteq}, \otimes \rangle$ turns out to be a unital, commutative quantale, where $\{\epsilon\} \in \wp(\mathtt{Sub})$ is the unit. In the following, we will slightly abuse the notation by applying the operation $\otimes$ also to substitutions.

### 4.1  A core logic programming language

We consider programs as (finite) sets of procedure declarations, and we assume that each procedure can be declared in at most one clause of the form $p(\bar{x}) \leftarrow A$. This assumption simplifies the treatment of condensing procedures: With each predicate $p$ a single clause is allowed in the program. The non-deterministic choice in the definition of $p$ is specified by allowing disjunction ($\sum$) in clause-bodies. The following syntax specifies the structure of logic programs considered in this section. In the following definition $\Theta \in \wp_{fin}(\mathtt{Sub})$ stands for a finite set of substitutions.

$$P ::= \varnothing \mid p(\bar{x}) \leftarrow A \mid P.P$$
$$A ::= \Theta \mid A \otimes A \mid \sum_{i=1}^{n} A_i \mid p(\bar{x})$$

The forward semantics $[\![\langle p(\bar{x}), \{\vartheta\}\rangle]\!]_P$ of a procedure call $p(\bar{x})\vartheta$ in a program $P$ is defined as $\mathcal{S}_{p(\bar{x})}(\{\vartheta\})$, as given by the following function on $\wp(\mathtt{Sub})$, which is recursively defined on program's structure for any $\Phi \in \wp(\mathtt{Sub})$:

$$
\begin{aligned}
\mathcal{S}_{\Theta}(\Phi) &= \Theta \otimes \Phi \\
\mathcal{S}_{A_1 \otimes A_2}(\Phi) &= \mathcal{S}_{A_1}(\Phi) \otimes \mathcal{S}_{A_2}(\Phi) \\
\mathcal{S}_{\sum_{i=1}^{n} A_i}(\Phi) &= \bigvee_{i=1}^{n} \mathcal{S}_{A_i}(\Phi) \\
\mathcal{S}_{p(\bar{x})}(\Phi) &= \mathcal{S}_A(\Phi) \quad \text{where } p(\bar{x}) \leftarrow A \lll P.
\end{aligned}
$$

In this definition $\lll$ selects a (renamed) clause from $P$ where variables not in $\bar{x}$ are renamed apart from $\bar{x}$ and $\Phi$. The forward concrete semantics of a logic program $P$ with initial goal $p(\bar{x})$ is therefore $F_{P,p(\bar{x})} = \lambda\Theta.[\![\langle p(\bar{x}), \Theta\rangle]\!]_P$.

Thus, the best correct approximation of $F_{P,p(\bar{x})}$ with respect to an abstract domain $\rho \in uco(\wp(\mathtt{Sub}))$ is inductively defined as follows for any $\Phi \in \rho(\wp(\mathtt{Sub}))$:

$$
\begin{aligned}
\mathcal{S}_{\Theta}^{\rho}(\Phi) &= \rho(\Theta \otimes \Phi) \\
\mathcal{S}_{A_1 \times A_2}^{\rho}(\Phi) &= \rho(\mathcal{S}_{A_1}^{\rho}(\Phi) \otimes \mathcal{S}_{A_2}^{\rho}(\Phi)) \\
\mathcal{S}_{\sum_{i=1}^{n} A_i}^{\rho}(\Phi) &= \rho(\bigvee_{i=1}^{n} \mathcal{S}_{A_i}^{\rho}(\Phi)) \\
\mathcal{S}_{p(\bar{x})}^{\rho}(\Phi) &= \mathcal{S}_A^{\rho}(\Phi) \quad \text{where } p(\bar{x}) \leftarrow A \lll P.
\end{aligned}
$$

As above, the abstract semantics of a procedure call $p(\bar{x})$ in a program $P$, with abstract initial call $\Theta \in \rho$, is defined as $[\![\langle p(\bar{x}), \Theta\rangle]\!]_P^{\rho} = \mathcal{S}_{p(\bar{x})}^{\rho}(\Theta)$. The forward abstract semantics of a logic program $P$ with initial goal $p(\bar{x})$ is therefore $F_{P,p(\bar{x})}^{\rho} = \lambda\Theta.[\![\langle p(\bar{x}), \Theta\rangle]\!]_P^{\rho}$. Note that, in each equation above, $\mathcal{S}^{\rho}$ is recursively defined as the best correct approximation of $\mathcal{S}$ in $\rho$.

### 4.2  Generalizing condensing domains

The first attempt to formally specify condensing procedures as an abstract domain property was due to Marriott and Søndergaard [1993]. The authors consider *downward-closed* abstract domains: $X \in \mathrm{uco}(\wp(\mathtt{Sub}))$ is downward-closed if any $\phi \in X$ is closed by instantiation. In this case, the glb of $X$, that is set intersection, actually plays the role of abstract unification. A domain $X \in \mathrm{uco}(\wp(\mathtt{Sub}))$ is called *condensing* if for any program $P$, query $Q$, and $\phi, \phi' \in X$, we have:

$$F_{P,Q}^X(\phi \wedge \phi') = \phi \wedge F_{P,Q}^X(\phi')$$

where $F_{P,Q}^X : X \longrightarrow X$ is the best correct approximation in $X$ of the goal-directed semantic function $F_{P,Q}$ mapping a set of initial substitutions for the program $P$ and query $Q$ to their semantics. Giacobazzi and Scozzari [1998] gave a characterization of condensing downward-closed abstract domains as so-called Heyting-closed domains. Heyting algebras are instances of quantales where linear implication is replaced by intuitionistic implication, i.e. the quantale multiplication is the meet operation. This perfectly models downward closed condensing abstract domains. Indeed, the collection of idempotent substitutions closed by instantiation, denoted by $\wp^{\downarrow}(\mathtt{Sub})$, is a complete Heyting algebra, i.e. a quantale $\langle \wp^{\downarrow}(\mathtt{Sub}), \wedge \rangle$ [Giacobazzi and Scozzari 1998]. In this section, we generalize this construction to any, possibly non downward-closed, abstract domain. This characterization relies upon the following generalized notion of condensing abstract domain, where we assume that $\langle \wp(\mathtt{Sub})_{\subseteq}, \otimes \rangle$ is a quantale.

*Definition* 4.1. An abstract domain $\rho \in \mathrm{uco}(\wp(\mathtt{Sub}))$ is *condensing* for $F_{P,Q}^\rho : \rho \longrightarrow \rho$, indexed on programs $P$ and queries $Q$, if for all $\Theta, \Phi \in \rho$,

$$F_{P,Q}^\rho(\rho(\Theta \otimes \Phi)) = \rho(\Theta \otimes F_{P,Q}^\rho(\Phi)).$$

This property depends upon the domain $\rho$ and the abstract semantics $F_{P,Q}^\rho$, which in turn is defined on $\rho$. Not all domains are condensing: Marriott and Søndergaard [1993] exhibit some non-condensing domains for groundness analysis. Let us see an example of an abstract domain which is not condensing.

*Example* 4.2. Two variables $x, y \in \mathcal{V}$ are said to be *independent* for the substitution $\theta$ when $vars(\theta(x)) \cap vars(\theta(y)) = \varnothing$. Let $I_{xy}$ be the set of substitutions for which $x$ and $y$ are independent:

$$I_{xy} \stackrel{\mathrm{def}}{=} \{\theta \in \mathtt{Sub} \mid vars(\theta(x)) \cap vars(\theta(y)) = \varnothing\}.$$

We consider a finite set of variables of interest $VI \subset_{\mathit{fin}} \mathcal{V}$, which are the relevant variables. According to this, abstract domains are restricted to have variables in $VI$ and do not explicitly show the set of relevant variables they refer to. The basic domain $\mathsf{PSh}_{VI}$ for detecting pair-sharing (i.e., pairs of variables which may share a common variable) is given by the most abstract domain which contains all the objects $I_{xy}$, for any $x, y \in VI$, with $x \neq y$:

$$\mathsf{PSh}_{VI} \stackrel{\mathrm{def}}{=} \mathcal{M}(\{I_{xy} \mid x, y \in VI, x \neq y\}).$$

The domain $\mathsf{PSh}_{VI}$ induces a Galois insertion $(\alpha, \wp(\mathtt{Sub}), \mathsf{PSh}_{VI}, \gamma)$ defined as follows: for all $\Theta \in \wp(\mathtt{Sub})$,

$$\alpha(\Theta) \stackrel{\mathrm{def}}{=} \bigwedge \{I_{xy} \mid x, y \in VI, x \neq y, \forall \theta \in \Theta \; vars(\theta(x)) \cap vars(\theta(y)) = \varnothing\}.$$

Let $P$ be the following program:

$$p(X, Y) \leftarrow \{\{X/a\}, \{Y/a\}\}.$$

For $VI = \{X, Y\}$ we have that $\mathsf{PSh}_{VI} = \{\top, I_{XY}\}$, where $\gamma(\top) = \mathtt{Sub}$. Let us denote by $\rho$ the uco associated with $\mathsf{PSh}_{VI}$. Note that the abstract operation $\otimes$ on $\mathsf{PSh}_{VI}$ is trivially defined as follows: For all $A, B \in \mathsf{PSh}_{VI}$, $\rho(A \otimes B) = \top$ (this is a consequence of the fact that $\{X/Z\} \otimes \{Y/Z\} = \{X/Z, Y/Z\}$ which does not belong to $I_{XY}$ while both $\{X/Z\}$ and $\{Y/Z\}$ do). If we compute $F^\rho_{P,p(X,Y)}$ with initial query $\Phi = \top$ we obtain:

$$
\begin{aligned}
F^\rho_{P,p(X,Y)}(\top) &= \mathcal{S}^\rho_{p(X,Y)}(\top) \\
&= \mathcal{S}^\rho_{\{\{X/a\},\{Y/a\}\}}(\top) \\
&= \rho(\{\{X/a\},\{Y/a\}\} \otimes \top) = I_{XY}.
\end{aligned}
$$

Therefore, being $\rho(I_{XY} \otimes \top) = \top$, the following results hold:

$$
F^\rho_{P,p(X,Y)}(\rho(I_{XY} \otimes \top)) = F^\rho_{P,p(X,Y)}(\top) = I_{XY}
$$

$$
\rho(I_{XY} \otimes F^\rho_{P,p(X,Y)}(\top)) = \rho(I_{XY} \otimes I_{XY}) = \top.
$$

As a consequence, the domain $\rho$ is not condensing for $F^\rho_{P,p(X,Y)}$. □

In the following we give a systematic method for designing condensing abstract domains for logic programs. This allows us to remove the possible loss of precision between goal-independent vs. goal-directed static program analyses. This is achieved by observing that $\{\epsilon\}$, which is the unit element in $\langle \wp(\mathtt{Sub})_\subseteq, \otimes \rangle$, represents the weakest possible substitution. Therefore, if a condensing abstract domain $\langle \rho(\wp(\mathtt{Sub}))_\subseteq, \rho \circ \otimes \rangle$ satisfying Definition 4.1, is also a commutative quantale with unit $\rho(\{\epsilon\})$, then for any set of substitutions $\Theta \in \rho(\wp(\mathtt{Sub}))$, we have that $\rho(\Theta \otimes F^\rho_{P,Q}(\rho(\{\epsilon\}))) = F^\rho_{P,Q}(\Theta)$. The idea here is that by computing the abstract semantics of a query with initial call in $\Theta$, i.e., $F^\rho_{P,Q}(\Theta)$, we obtain the same result as unifying the substitutions in $\Theta$ with the result of the semantics of the same query in the most general environment $\epsilon$, which is $F^\rho_{P,Q}(\rho(\{\epsilon\}))$. This encodes the typical way we derive the analysis of a query in an initial state from a goal-independent (condensing) analysis: We filter out, by unification, those computations which do not satisfy the given initial state (cf. [Barbuti et al. 1993]).

### 4.3 Weak completeness

It is worth noting that the abstract semantics of a program $P$ is always defined by iterating abstract unification of a concrete substitution belonging to $P$ against the result of the previous computation, which is an abstract object. The fixpoint of this iterated procedure gives the semantics of a predicate. As observed above, an abstract domain is condensing when it is possible to postpone the evaluation of a specific call after the evaluation of the semantics of each predicate, without any loss of precision. This means that it is possible to propagate the information contained in a query back to the semantics without recomputing the semantics of that query, by a simple unification operation. In this case, the semantics is obtained by computing the semantics of each predicate with the most general call. It is clear that completeness is sufficient to ensure condensation, since all intermediate abstractions can be removed from the fixpoint computation of the semantics of each predicate. However, a weaker form of completeness can be considered in view of the evaluation strategy implemented in the bottom-up semantics of logic programs [Barbuti et al. 1993; Codish et al. 1994]. The idea is that only one of the two arguments of unification is an abstract object. Our aim here is to formalize the intuition that still no loss of precision is accumulated in the abstract computation when at least one argument of $\otimes$ is an abstract object. This happens in logic program analysis, when $\otimes$ is unification and the semantics consists in iteratively unifying concrete and abstract objects, which respectively come from program's clauses and the current abstract substitutions. We define an abstract domain $\rho \in \text{uco}(C)$ to be *weak-complete* for $\otimes$ when:

$$\text{if either } x \in \rho \text{ or } y \in \rho \text{ then } \rho(\rho(x) \otimes \rho(y)) = \rho(x \otimes y).$$

Hence, this is equivalent to require that $\rho$ satisfies the following equalities:

$$\rho \circ \otimes \circ \langle \rho, \rho \rangle = \rho \circ \otimes \circ \langle \rho, id \rangle = \rho \circ \otimes \circ \langle id, \rho \rangle.$$

In turn, by the hypothesis of commutativity of $\otimes$, this last condition is equivalent to the following single equation:

$$\rho \circ \otimes \circ \langle \rho, \rho \rangle = \rho \circ \otimes \circ \langle \rho, id \rangle. \tag{1}$$

It is worth pointing out that this is actually a weakening of standard completeness, i.e., any $\rho$ complete for $\otimes$ is weak-complete for $\otimes$ as well. The converse does not hold.

Then, for a given abstract domain $A \in \text{uco}(C)$, we are interested in characterizing the most abstract domain $\rho \in \text{uco}(C)$ which is more concrete than $A$ and satisfies Equation (1). This domain, when it exists, is called the *weak-complete shell* of $A$ for $\otimes$. Weak-completeness problems can be solved by exploiting the same technique used for completeness, i.e., by resorting to a recursive abstract domain equation involving linear implication. The next theorem gives a recursive characterization of the solutions of Equation (1).

THEOREM 4.3. *Let $\langle C_{\le}, \otimes \rangle$ be a unital, commutative quantale and $\rho \in \text{uco}(C)$. The following are equivalent.*

(i)  $\rho \circ \otimes \circ \langle \rho, \rho \rangle = \rho \circ \otimes \circ \langle \rho, id \rangle$;

(ii)  $\rho$ is complete for $F_\rho = \{\lambda y.x \otimes y \mid x \in \rho\}$;

(iii)  $\rho = \rho \sqcap (\rho \overset{\wedge}{\multimap} \rho)$.

PROOF. $\rho \circ \otimes \circ \langle \rho, \rho \rangle = \rho \circ \otimes \circ \langle \rho, id \rangle$ holds iff for all $x \in \rho$ and $y \in C$ it holds $\rho(x \otimes \rho(y)) = \rho(x \otimes y)$, that is to say that $\rho$ is complete for the set of unary functions $F_\rho = \{\lambda y.x \otimes y \mid x \in \rho\}$. By Theorem 2.1, $\rho$ is complete for $F_\rho$ iff $\rho \sqsubseteq \mathcal{R}_{F_\rho}(\rho)$. By Theorem 3.3, this is equivalent to say that $\rho \sqsubseteq \rho \overset{\wedge}{\multimap} \rho$, and therefore $\rho = \rho \sqcap \rho \overset{\wedge}{\multimap} \rho$. $\square$

COROLLARY 4.4. *Let $\langle C_{\le}, \otimes \rangle$ be a unital, commutative quantale and $A \in \text{uco}(C)$. The weak-complete shell of $A$ for $\otimes$ is $gfp(\lambda X.A \sqcap X \sqcap (X \overset{\wedge}{\multimap} X))$.*

PROOF. Since the operator $\lambda X.X \overset{\wedge}{\multimap} X$ is clearly monotone[1], from Theorem 4.3 it directly follows that the most abstract domain which includes $A$ and is weak-complete for $\otimes$ is given by $gfp(\lambda X.A \sqcap X \sqcap X \overset{\wedge}{\multimap} X)$. $\square$

---

[1]It is worth noting that, even if $\mathcal{R}_{F_\rho} = \lambda \eta.\rho \overset{\wedge}{\multimap} \eta$ is co-additive for any $\rho \in \text{uco}(C)$, this does not imply that the operator $\lambda \eta.\eta \overset{\wedge}{\multimap} \eta$ is co-additive as well. This is a consequence of the fact that the set of functions $F_\rho$ for which we want to be complete, changes at each iteration.

Thus, the weak-complete shell of a domain $A \in \mathrm{uco}(C)$ is exactly the greatest solution in $\mathrm{uco}(C)$ of the following recursive abstract domain equation:

$$X = A \sqcap X \sqcap (X \stackrel{\curvearrowright}{\multimap} X). \tag{2}$$

### 4.4 Condensing domains in logical form

We are now in the position to prove the main result of this section. In the following we prove that any abstract domain, which is solution of the recursive domain equation (1), is a unital commutative quantale which satisfies the relevant property of being *condensing* and, under additional non-restrictive hypotheses, condensing domains are all and only the solutions of Equation (1). This result shows a surprising link between completeness in abstract interpretation and condensing domains and, more importantly, it gives computational relevance in static program analysis to the notion of weak-completeness. Before proving this result, let us show the following simple property which will be used later.

LEMMA 4.5. *Let $\langle C_{\leq}, \otimes \rangle$ be a unital, commutative quantale, $\rho \in \mathrm{uco}(C)$ and $c_1, .., c_n \in C$ for $n > 0$. If $\rho$ satisfies Eq. (1), then for all $i$, with $1 \leq i \leq n$, it holds: $\rho(\rho(c_1) \otimes \ldots \otimes \rho(c_{n-1}) \otimes \rho(c_n)) = \rho(\rho(c_1) \otimes \ldots \otimes \rho(c_{n-1}) \otimes c_n)$.*

PROOF. The proof is by induction on the number of applications of $\otimes$.
($n = 0$) is straightforward: $\rho(\rho(c)) = \rho(c)$.
($n > 1$) follows by Eq. (1):

$$
\begin{aligned}
\rho(\rho(c_1) \otimes \rho(c_2) \otimes \ldots \otimes \rho(c_{n-1}) \otimes \rho(c_n)) &= \text{[by Eq. (1)]} \\
\rho(\rho(c_1) \otimes \rho(\rho(c_2) \otimes \ldots \otimes \rho(c_{n-1}) \otimes \rho(c_n))) &= \text{[by inductive hypothesis]} \\
\rho(\rho(c_1) \otimes \rho(\rho(c_2) \otimes \ldots \otimes \rho(c_{n-1}) \otimes c_n)) &= \text{[by Eq. (1)]} \\
\rho(\rho(c_1) \otimes \rho(c_2) \otimes \ldots \otimes \rho(c_{n-1}) \otimes c_n). &\quad \square
\end{aligned}
$$

THEOREM 4.6. *Let $P$ be a program and $\rho \in \mathrm{uco}(\wp(\mathtt{Sub}))$. If $\rho = \rho \sqcap \rho \stackrel{\curvearrowright}{\multimap} \rho$ then $\rho$ is condensing for $F^{\rho}_{P, p(\bar{x})}$.*

PROOF. Let $\rho$ be a solution of the recursive domain equation $\rho = \rho \sqcap \rho \stackrel{\curvearrowright}{\multimap} \rho$. Let $\Psi, \Theta \in \rho$. Since $\langle \wp(\mathtt{Sub})_{\subseteq}, \otimes \rangle$ is a unital, commutative quantale, it is sufficient to prove that $\mathcal{S}^{\rho}$ is condensing, i.e., $\mathcal{S}^{\rho}_A(\rho(\Psi \otimes \Theta)) = \rho(\Psi \otimes \mathcal{S}^{\rho}_A(\Theta))$, for any procedure definition $A$. This is proved by induction on the structure of the procedure definition $A$. Let $\Phi, \Psi, \Theta \in \rho(\wp(\mathtt{Sub}))$.

—Consider $\mathcal{S}^{\rho}_{\Phi}$. By Lemma 4.5 we have that

$$
\begin{aligned}
\mathcal{S}^{\rho}_{\Phi}(\rho(\Psi \otimes \Theta)) &= \rho(\Phi \otimes \rho(\Psi \otimes \Theta)) \\
&= \rho(\Phi \otimes \Psi \otimes \Theta) \\
&= \rho(\Psi \otimes \Phi \otimes \Theta) \\
&= \rho(\Psi \otimes \rho(\Phi \otimes \Theta)) \\
&= \rho(\Psi \otimes \mathcal{S}^{\rho}_{\Phi}(\rho(\Theta))).
\end{aligned}
$$

—Consider $\mathcal{S}^{\rho}_{A_1 \wedge A_2}$. By Lemma 4.5 and inductive hypothesis we have that

$$
\begin{aligned}
\mathcal{S}^{\rho}_{A_1 \wedge A_2}(\rho(\Psi \otimes \Theta)) &= \rho(\mathcal{S}^{\rho}_{A_1}(\rho(\Psi \otimes \Theta)) \otimes \mathcal{S}^{\rho}_{A_2}(\rho(\Psi \otimes \Theta))) \\
&= \rho(\Psi \otimes \mathcal{S}^{\rho}_{A_1}(\Theta) \otimes \Psi \otimes \mathcal{S}^{\rho}_{A_2}(\Theta)) \\
&= \rho(\Psi \otimes \rho(\mathcal{S}^{\rho}_{A_1}(\Theta) \otimes \mathcal{S}^{\rho}_{A_2}(\Theta))) \\
&= \rho(\Psi \otimes \mathcal{S}^{\rho}_{A_1 \wedge A_2}(\Theta)).
\end{aligned}
$$

—Consider $\mathcal{S}^{\rho}_{\sum_{i=1}^n A_i}$. Recall that, by definition of quantale, the operation $\otimes$ is additive. Therefore, by inductive hypothesis and Lemma 4.5 we have that

$$
\begin{aligned}
\mathcal{S}^{\rho}_{\sum_{i=1}^n A_i}(\Psi \otimes \Theta) &= \rho(\textstyle\bigvee_{i=1}^n \mathcal{S}^{\rho}_{A_i}(\Psi \otimes \Theta)) \\
&= \rho(\textstyle\bigvee_{i=1}^n \Psi \otimes \mathcal{S}^{\rho}_{A_i}(\Theta)) \\
&= \rho(\Psi \otimes \textstyle\bigvee_{i=1}^n \mathcal{S}^{\rho}_{A_i}(\Theta)) \\
&= \rho(\Psi \otimes \rho(\textstyle\bigvee_{i=1}^n \mathcal{S}^{\rho}_{A_i}(\Theta))) \\
&= \rho(\Psi \otimes \mathcal{S}^{\rho}_{\sum_{i=1}^n A_i}(\Theta)).
\end{aligned}
$$

—Consider $\mathcal{S}^{\rho}_{p(\bar{x})}$. The thesis follows immediately by definition and inductive hypothesis. $\square$

The key point in the previous theorem is that the operation $\otimes$ is additive, i.e., it distributes over arbitrary lub's. Thus, whenever the abstract domain contains all linear implications between any two abstract objects, no loss of precision is accumulated by distributing the abstract unification over non-deterministic computations, which are modeled by abstract disjunction.

In order to prove the converse of Theorem 4.6, i.e., that condensing domains are solutions of Eq. (1), we need some additional hypotheses. The next result, together with Theorem 4.6, provides a constructive characterization of condensing domains which are at least as precise as a given domain $A$. These are all and only those domains which are solutions of a recursive domain equation of the form $X = A \sqcap X \sqcap X \overset{\wedge}{\multimap} X$.

THEOREM 4.7. *Let $P$ be a program, $p(\bar{x})$ be an atom, $\rho \in \mathrm{uco}(\wp(\mathtt{Sub}))$ and let $X$ be a generator for $\rho$, i.e., $\mathcal{M}(X) = \rho$. Assume that for any $\Phi, \Psi \in X$ there exist $\{\Theta_1, \ldots, \Theta_n\} \subseteq \mathtt{Sub}$, with $n > 0$, such that $\Theta_i$ is finite and the following conditions hold:*

—$\bigvee_{i=1}^n \Theta_i \subseteq \Phi \multimap \Psi$

—$\rho(\bigvee_{i=1}^n \Theta_i) = \rho(\Phi \multimap \Psi)$

*If $\rho$ is condensing for $\lambda\Theta \in \rho.[\![\langle p(\bar{x}), \Theta\rangle]\!]^{\rho}_P$ then $\rho = \rho \sqcap \rho \overset{\wedge}{\multimap} \rho$.*

PROOF. Let $\rho$ be condensing. We have to prove that $\rho \sqsubseteq \rho \overset{\wedge}{\multimap} \rho$, i.e., that for any $\Phi, \Psi \in \rho$, $\rho(\Phi \multimap \Psi) = \Phi \multimap \Psi$. Since $X$ is a generator set for $\rho$, it suffices to show that for any $\Phi, \Psi \in X$, $\rho(\Phi \multimap \Psi) = \Phi \multimap \Psi$. Suppose, by contradiction, that there exist $\Phi, \Psi \in X$ such that $\rho(\Phi \multimap \Psi) \supset \Phi \multimap \Psi$. Note that, by definition, $\Phi \multimap \Psi$ is the most abstract object $\Delta$ which satisfies $\Phi \otimes \Delta \subseteq \Psi$. Therefore, $\Phi \otimes \rho(\Phi \multimap \Psi) \not\subseteq \Psi$. By hypothesis there exists $\{\Theta_1, \ldots \Theta_n\} \subseteq \mathtt{Sub}$, with $n > 0$, such that $\Theta_i$ is finite and the above conditions hold. Consider the program $P = \{p :\!- \sum_{i=1}^n \Theta_i\}$ consisting of $n$ facts. Then, by hypothesis, we have that:

$$
\begin{aligned}
\rho(\Phi \otimes [\![\langle p, \{\epsilon\}\rangle]\!]^\rho_P) &= \rho(\Phi \otimes \rho(\bigvee_{i=1}^n \rho(\{\epsilon\} \otimes \Theta_i))) \\
&= \rho(\Phi \otimes \rho(\bigvee_{i=1}^n \rho(\Theta_i))) \\
&= \rho(\Phi \otimes \rho(\bigvee_{i=1}^n \Theta_i)) \\
&= \rho(\Phi \otimes \rho(\Phi \multimap \Psi)) \\
&\not\sqsubseteq \Psi \\
&\sqsupseteq \rho(\Phi \otimes (\Phi \multimap \Psi)) \\
&\sqsupseteq \rho(\Phi \otimes \bigvee_{i=1}^n \Theta_i) \\
&= \rho(\bigvee_{i=1}^n \Phi \otimes \Theta_i) \\
&= \rho(\bigvee_{i=1}^n \rho(\Phi \otimes \Theta_i)) \\
&= [\![\langle p, \Phi\rangle]\!]^\rho_P \\
&= [\![\langle p, \rho(\Phi \otimes \{\epsilon\})\rangle]\!]^\rho_P
\end{aligned}
$$

Therefore $\rho$ would not be condensing, which is a contradiction.   $\square$

The hypothesis in Theorem 4.7 is not restrictive for most domains used in logic program analysis. Conditions (1) and (2) say that any implicational object $x \multimap y$ always approximates a finite disjunction of substitutions. This allows us, in Theorem 4.7, to construct, for any implicational object $x \multimap y$, a (finite) program having that object in the abstract semantics. The idea is that, in order to prove the converse of Theorem 4.6, any implicational object has to be the semantics of some well-defined program. The following characterization of condensing is therefore immediate by Theorems 4.6 and 4.7.

COROLLARY 4.8. *Under the hypothesis of Theorem 4.7, $\rho$ is condensing for the semantic function $\lambda\Theta \in \rho.[\![\langle p(\bar{x}), \Theta\rangle]\!]^\rho_P$ if and only if $\rho = \rho \sqcap \rho \overset{\wedge}{\multimap} \rho$.*

*Example* 4.9. Consider the domain $\rho = \{\top, I_{XY}\}$ as defined in Example 4.2. The following equivalences hold.

—for all $\Theta \in \wp(\mathtt{Sub})$, from the definition of $\multimap$ it follows that $\Theta \multimap \top = \top$.

—$\top \multimap I_{XY} = \{\theta \in \mathtt{Sub} \mid \forall \delta \in \mathtt{Sub} \; \theta \otimes \delta \sqsubseteq I_{XY}\} = \{\theta \in \mathtt{Sub} \mid \; \downarrow \theta \subseteq I_{XY}\}$. If either $X$ or $Y$ is ground in $\theta$, then the result immediately follows. Otherwise, it is always possible to find an instance of $\theta$ where $X$ and $Y$ share a common variable. Thus we have that $\top \multimap I_{XY} = \{\theta \in \mathtt{Sub} \mid vars(\theta(X)) = \emptyset \text{ or } vars(\theta(Y)) = \emptyset\}$.

—$I_{XY} \multimap I_{XY} = \{\theta \in \mathtt{Sub} \mid \forall \delta \in I_{XY} \; \theta \otimes \delta \sqsubseteq I_{XY}\}$. If either $X$ or $Y$ is ground in $\theta$, then it trivially holds $\theta \otimes \delta \sqsubseteq I_{XY}$. Consider now the case that both $X$ and $Y$ are not ground. Recall that, given a substitution $\delta \in I_{XY}$, all variables but $X$ and $Y$ are allowed to share. Therefore, when unifying $\theta \otimes \delta$, we have to assure that no variable in $\theta$ shares with any other variable. For example, if we consider a substitution $\theta = \{Z/W\}$, then by unifying with $\delta = \{X/Z, Y/W\}$ we immediately obtain a substitution which does not belong to $I_{XY}$. As a consequence, $I_{XY} \multimap I_{XY} = \{\theta \in \mathtt{Sub} \mid vars(\theta(X)) = \emptyset \text{ or } vars(\theta(Y)) = \emptyset\} \cup \{\theta \in \mathtt{Sub} \mid \forall v \in dom(\theta) \; vars(\theta(v)) = \emptyset\}$.

Let us denote by $G_{XY}$ the set $\{\theta \in \mathtt{Sub} \mid vars(\theta(X)) = \emptyset \text{ or } vars(\theta(Y)) = \emptyset\}$ and by $\epsilon_G$ the set $\{\theta \in \mathtt{Sub} \mid \forall v \in dom(\theta) \; vars(\theta(v)) = \emptyset\}$. Since $G_{XY} \subseteq G_{XY} \cup \epsilon_G \subseteq I_{XY} \subseteq \top$, we have that $\rho' = \rho \sqcap \rho \overset{\wedge}{\multimap} \rho = \{G_{XY}, G_{XY} \cup \epsilon_G, I_{XY}, \top\}$. It is now easily seen that $\rho' \overset{\wedge}{\multimap} \rho' = \rho'$ and therefore the most abstract solution to the domain equation $\rho = \rho \sqcap \rho \overset{\wedge}{\multimap} \rho$ is $\rho' = \rho \cup \{G_{XY}, G_{XY} \cup \epsilon_G\}$. By using $\rho'$ we now obtain:

$$F_{P,p(X,Y)}^{\rho'}(\rho'(I_{XY} \otimes \top)) = F_{P,p(X,Y)}^{\rho'}(\top) = \rho'(\{\{X/a\}, \{Y/a\}\} \otimes \top) = G_{XY}$$

and

$$\rho'(I_{XY} \otimes F_{P,p(X,Y)}^{\rho'}(\top)) = \rho'(I_{XY} \otimes G_{XY}) = G_{XY}.$$

## 5. CONCLUSION

We have shown a surprising link between completeness in quantale-like structures and condensation. This provides a characterization of condensing domains as models of a fragment of propositional linear logic. The relation between completeness and reversible dataflow analysis has gained great attention in the last few years. As observed in [King and Lu 2002], the possibility of reusing code in logic programming is often related to the problem of figuring out how to query a program, and backward analysis allows us to automatically derive the possible modes in which predicates must be called. As proved in [King and Lu 2002], this property needs condensing abstract domains. By this observation and from our characterization of condensing abstract domains in logical form, it seems possible to characterize reversible abstract interpretations in a pure domain-theoretic form. There are still many open questions along this line of research. It is for instance a major challenge to design condensing abstract domains for aliasing. Theorems 4.6 and 4.7 give necessary and sufficient conditions to systematically design these domains, but the construction of non-downward closed condensing abstract domains, although clarified and made systematic, is still quite difficult due to the complex structure of the quantale of idempotent substitutions. This is the case if we are looking for the most abstract condensing domain refining *sharing*, which is an abstract domain devoted to the static analysis of variable aliasing in idempotent substitutions [Jacobs and Langen 1992; Langen 1990]. In this case, the solution of the abstract domain equation $X = sharing \sqcap X \sqcap (X \overset{\wedge}{\multimap} X)$ is still unknown. Our results can be used also to prove that known domains are condensing. Scozzari [2002] proved that the domain *Pos* for groundness analysis [Armstrong et al. 1998] is the most abstract solution of the abstract domain equation $X = \mathcal{G} \sqcap X \rightarrow X$, where $\mathcal{G}$ is the domain of plain groundness in [Jones and Søndergaard 1987]. In view of Theorems 4.6 and 4.7, this provides an alternative proof of the known fact that *Pos* is condensing [Marriott and Søndergaard 1993]. The advantage of our method with respect to other proofs is that it gives a constructive procedure to systematically design condensing domains even for non-downward closed properties.